\documentclass[aps,prd,twocolumn,superscriptaddress,nofootinbib]{revtex4}

\usepackage{latexsym}
\usepackage{amsmath}
\usepackage{amssymb}
\usepackage{amsfonts}
\usepackage{slashed}

\usepackage{color}

\usepackage{supertabular} 
\usepackage{placeins}
\usepackage{epsfig}
\usepackage{graphicx}

\usepackage{color,soul}

\usepackage[breaklinks,pageanchor]{hyperref}

\hypersetup{
  colorlinks=true,
  linkcolor=blue,
  filecolor=magenta,
  urlcolor=blue,
  citecolor=blue,
}

\begin{document}


\title{Spectroscopy of $\mathbf{B_c}$ mesons and the possibility of finding exotic $\mathbf{B_c}$-like structures}

\author{Pablo G. Ortega}
\email[]{pgortega@usal.es}
\affiliation{Grupo de F\'isica Nuclear and Instituto Universitario de F\'isica Fundamental y Matem\'aticas (IUFFyM), Universidad de Salamanca, E-37008 Salamanca, Spain}

\author{Jorge Segovia}
\email[]{jsegovia@upo.es}
\affiliation{Departamento de Sistemas F\'isicos, Qu\'imicos y Naturales, \\ Universidad Pablo de Olavide, E-41013 Sevilla, Spain}

\author{David R. Entem}
\email[]{entem@usal.es}
\affiliation{Grupo de F\'isica Nuclear and Instituto Universitario de F\'isica Fundamental y Matem\'aticas (IUFFyM), Universidad de Salamanca, E-37008 Salamanca, Spain}

\author{Francisco Fern\'andez}
\email[]{fdz@usal.es}
\affiliation{Grupo de F\'isica Nuclear and Instituto Universitario de F\'isica 
Fundamental y Matem\'aticas (IUFFyM), Universidad de Salamanca, E-37008 
Salamanca, Spain}

\date{\today}

\begin{abstract}
The bottom-charmed ($B_c$) mesons are more stable than their charmonium ($c\bar c$) and bottomium ($b\bar b$) partners because they cannot annihilate into gluons. However, the low production cross-sections and signal-to-background ratios avoided until now their clear identification. The recent experimental results reported by CMS and LHCb at CERN open the possibility of having a $B_c$ spectrum as complete as the ones of charmonium and bottomonium. Motivated by this expectation, we compute bottom-charmed meson masses in the region energies in which decay meson-meson thresholds are opened, looking for the analogs to the $X(3872)$ in the $B_c$ spectroscopy. We use a constituent quark model in which quark-antiquark degrees of freedom are complemented by four-body Fock states configurations. The model has been applied to a wide range of hadronic observables, in particular to the $X(3872)$, and thus the model parameters are completely constrained. No extra states are found in the $J^P=0^+$ and $J^P=1^+$ sectors. However, in the $J^P=2^+$ sector we found an additional state very close to the $D^*B^*$ threshold which could be experimentally detected.
\end{abstract}



\maketitle


\section{INTRODUCTION}
\label{sec:intro}

The bottom-charmed ($B_c$) meson family provides a unique window to test the non-relativistic limit of quantum chromodynamics (QCD), the strong interaction sector of the Standard Model of Particle Physics, because they are the only quarkonium bound-states consisting of heavy quarks with different flavors: either $c\bar b$ for the positive-charged channel or $b\bar c$ for the negative one. There is an extra reward on studying these open-flavor bound-state systems: contrary to charmonium ($c\bar c$) and bottomonium ($b\bar b$), the $B_c$ mesons cannot annihilate into gluons and thus these states are very stable, with narrow widths, at least for those which are below the lowest strong-decay $B^{(\ast)}D^{(\ast)}$-thresholds.  

The observation of the $B_c(1^1S_0)$ bound-state\footnote{The spectroscopic notation $n^{2S+1}L_J$ is used, where $n=1$ indicates the ground state and $n=2,3,\ldots$ the respective excited states with higher energies but equal $J^P$ (following the notation of PDG), $S$ the total spin of the two valence quarks, $L$ their relative angular momentum where $S,\,P,\,D,\,F\,\ldots$ implies, respectively, $L=0,\,1,\,2,\,3,\ldots$, and $J$ is the total angular momentum of the system.} by the CDF Collaboration at the Tevatron collider in $1998$~\cite{Abe:1998wi, Abe:1998fb} demonstrated the feasibility of studying experimentally the $B_c$ spectroscopy. However, there were not new signals of bottom-charmed mesons during almost twenty years because of low production cross-sections, large backgrounds and relatively-easy misidentifications. The ATLAS Collaboration~\cite{Aad:2014laa} reported in $2014$ the observation of a peak at $6842\pm4\pm5\,\text{MeV/c}^2$ which was interpreted as either the $B_c^\ast(2^3S_1)$ excited state or an unresolved pair of peaks from the decays $B_c(2^1S_0)\to B_c(1^1S_0)\pi^+\pi^-$ and $B_c^\ast(2^3S_1)\to B_c^\ast(1^3S_1)\pi^+\pi^-$ followed by $B_c(1^3S_1)\to B_c(1^1S_0)\gamma$. Five years later, the CMS~\cite{Sirunyan:2019osb} and LHCb~\cite{PhysRevLett.122.232001} Collaborations released signals consistent with the $B_c(2S)$ and $B_c^\ast(2S)$ states observed in the $B_c(1S)\pi^+\pi^-$ invariant mass spectrum. LHCb Collaboration ~\cite{PhysRevLett.122.232001} reported two peaks located at

\begin{align}
6841.2 \pm 0.6 (stat) \pm 0.1 (syst) \pm 0.8 (B_c^+) \, \text{MeV/c}^2 \,,\\
\nonumber\\
6872.1 \pm 1.3 (stat) \pm 0.1 (syst) \pm 0.8 (B_c^+) \, \text{MeV/c}^2 \,,
\end{align}
which were assigned to the $B_c^\ast(2S)$ and the $B_c(2S)$ states respectively. CMS~\cite{Sirunyan:2019osb} observed two well-resolved peaks but only assigned a mass of $6871\pm 1.2 (stat) \pm 0.8 (syst) \pm 0.8 (B_c)$ MeV (where the last term is the uncertainty  in the world-average $B_c$ mass), to the $B_c(2S)$ state.
Contrary to theoretical expectations, the peak of the $B_c^\ast(2S)$ appears lower in energy than the $B_c(2S)$ due to the unresolved photon energy in the decay $B_c^\ast \to B_c \gamma$.
More results on $B_c$ mesons are expected to be reported in the near future and the scientific community is eager to analyze them.

On the theoretical side, non-relativistic quark models have been successfully applied to charmonium and bottomonium systems. The spectrum of $B_c$ mesons provides another opportunity to test them since the $B_c$ family shares dynamical properties with both the $c\bar c$ and $b\bar b$ sectors. Moreover, their results~\cite{Kwong:1990am, Eichten:1994gt, Kiselev:1994rc, Eichten:2019gig, Fulcher:1998ka,Li:2019tbn,Soni:2017wvy} can be contrasted with those obtained by relativistic approaches~\cite{Godfrey:1985xj, Zeng:1994vj, Gupta:1995ps, Ebert:2002pp, Ikhdair:2003ry, Godfrey:2004ya, Ikhdair:2004hg, Ikhdair:2004tj}, continuum functional methods for QCD~\cite{Chen:2020ecu,Chang:2019wpt,Yin:2019bxe}, QCD sum rules~\cite{Kiselev:1994rc,Wang:2012kw,Wang:2013cha}, effective field theories~\cite{Brambilla:2000db, Penin:2004xi, Peset:2018ria, Peset:2018jkf}, lattice-QCD~\cite{Allison:2004be,Mathur:2018epb,Dowdall:2012ab}. A collection of all should provide a reliable template from which compare the future experimental findings. In fact, there is some global agreement about which conventional $B_c$ states must exists below the $B^{(\ast)}D^{(\ast)}$-thresholds: there should be two sets of $S$-wave states, as many as two $P$-wave multiplets (the $1P$ and some or all of the $2P$), one $D$-wave multiplet below $BD$ threshold, and the $F$-wave multiplet should be sufficiently close to threshold that they may also be relatively narrow due to angular momentum barrier suppression of the Okubo-Zweig-Iizuka
(OZI)-rule~\cite{Okubo:1963fa, zweigcern2, iizuka1966systematics}.


The complications with the $B_c$ spectroscopy could begin at the energy region in which strong-decay meson-meson thresholds could play an important role in the formation of $B_c$(-like) structures. This has been vigorously manifested in the heavy quarkonium spectrum with the discovery of almost two dozen of charmonium- and bottomonium-like $XYZ$ states, which have forced the end of an era when heavy quarkonium was considered as a relatively well 
established heavy quark-antiquark bound-state system (see, e.g., 
reviews~\cite{Brambilla:2010cs, Brambilla:2014jmp, Olsen:2014qna} for more 
details on the experimental and theoretical situation on the subject).

The $X(3872)$, firstly discovered by Belle~\cite{Choi:2003ue} and sooner confirmed by CDF~\cite{Acosta:2003zx}, D0~\cite{Abazov:2004kp} and BaBar~\cite{Aubert:2004ns}, is the most prominent example of a charmonium-like structure whose closeness to the $D^0D^{\ast 0}$ threshold and its decay properties resemble an exotic composition; in particular, a $DD^\ast$ molecule with a possible $J^{PC}=1^{++}$ $c\bar c$ component manifesting at short distances. The $XYZ$ puzzlement has revived the old idea of existing deuteron-like states in the charmonium spectrum~\cite{DeRujula:1976zlg, Bander:1975fb} and the concept of meson-meson molecule has regained attention~\cite{Tornqvist:1993ng, Tornqvist:2004qy, Close:2003sg, Voloshin:2003nt}. 

The molecular picture leads to an immediate logical consequence: once a molecule is unequivocally determined, one can use QCD approximate symmetries to establish other meson-meson bound-states in other channels and sectors~\cite{Tornqvist:1993ng, Nieves:2012tt, Guo:2013sya}. For instance, if the interaction is assumed to be practically independent on the mass of the heavy quark (antiquark), molecules detected in the charmonium sector are expected to be reproduced in the bottomonium and bottom-charmed sectors with even larger binding energy, due to the reduction of the kinetic energy by the larger mass of the $b$ quark~\cite{Guo:2013sya}. 

It is important to remark herein that nearby quark-antiquark states can mix with the molecular ones and, then, change their composition, binding energy and decay properties in such a way that this effect must be taken into account when exploring the possible analogs of the $X(3872)$ in other heavy quark sectors. This has been done in Refs.~\cite{Nieves:2012tt, Entem:2016ojz, Cincioglu:2016fkm} for the charmonium case and in Refs.~\cite{Nieves:2012tt, Entem:2016ojz, Liu:2019stu} for the bottomonium one. This approach, in which we expand Fock's space to include, together with the degrees of freedom of two quarks, states of four quark, is different from the so-called tetraquark approach, where pure four-quark bound states are sought~\cite{Esposito:2016noz,PhysRevLett.119.202002,Luo:2017eub,Agaev:2017uky}.

In this manuscript we explore analogs of the $X(3872)$ state in the $B_c$ spectrum. In order to do this, we use a non-relativistic constituent quark model~\cite{Vijande:2004he} in which quark-antiquark and meson-meson degrees of freedom are incorporated (see references~\cite{Segovia:2013wma} and~\cite{Ortega:2012rs} for reviews). This model has been successfully applied to the charmonium and bottomonium sectors, studying their spectra~\cite{Segovia:2008zz, Segovia:2010zzb, Segovia:2016xqb}, their electromagnetic, weak and strong decays and reactions~\cite{Segovia:2011zza, Segovia:2012cd, Segovia:2013kg, Segovia:2014mca, Segovia:2011tb}, their coupling with meson-meson thresholds~\cite{Ortega:2009hj, Ortega:2016hde, Ortega:2017qmg, Ortega:2018cnm} and, lately, the phenomenological exploration of multiquark structures~\cite{Vijande:2006jf, Yang:2015bmv, Yang:2018oqd}. 

The manuscript is arranged as follows. In Sec.~\ref{sec:theory} we describe the main properties of our theoretical formalism giving details about the approaches used to describe the quark-antiquark sector, the meson-meson sector and the coupling between them. Section~\ref{sec:results} is devoted to present our results for the $B_c$ analogs of the $X(3872)$ state. We finish summarizing and giving some conclusions in Sec.~\ref{sec:epilogue}. 

\section{THEORETICAL FORMALISM}
\label{sec:theory} 

\subsection{Naive quark model}
\label{subsec:quarkmodel}

Constituent light quark masses and Goldstone-boson exchanges, which are consequences of dynamical chiral symmetry breaking in QCD, together with the perturbative one-gluon exchange (OGE) and a nonperturbative confining interaction are the main pieces of our constituent quark model~\cite{Vijande:2004he, Segovia:2013wma}.

A simple Lagrangian invariant under chiral transformations can be written in the following form~\cite{Diakonov:2002fq}
\begin{equation}
{\mathcal L} = \bar{\psi}(i\, {\slash\!\!\! \partial} - M(q^{2}) U^{\gamma_{5}}) \,\psi \,,
\end{equation}
where $M(q^2)$ is the dynamical (constituent) quark mass and $U^{\gamma_5} = e^{i\lambda _{a}\phi ^{a}\gamma _{5}/f_{\pi}}$ is the matrix of Goldstone-boson fields that can be expanded as
\begin{equation}
U^{\gamma _{5}} = 1 + \frac{i}{f_{\pi}} \gamma^{5} \lambda^{a} \pi^{a} - \frac{1}{2f_{\pi}^{2}} \pi^{a} \pi^{a} + \ldots
\end{equation}
The first term of the expansion generates the constituent quark mass while the second gives rise to a one-boson exchange interaction between quarks. The main contribution of the third term comes from the two-pion exchange which has been simulated by means of a scalar-meson exchange potential.

In the heavy quark sector chiral symmetry is explicitly broken and Goldstone-boson exchanges do not appear. However, it constrains the model parameters through the light-meson phenomenology~\cite{Segovia:2008zza} and provides a natural way to incorporate the pion exchange interaction in the molecular dynamics.

It is well known that multi-gluon exchanges produce an attractive linearly rising potential proportional to the distance between infinite-heavy quarks. However, sea quarks are also important ingredients of the strong interaction dynamics that contribute to the screening of the rising potential at low momenta and eventually to the breaking of the quark-antiquark binding string~\cite{Bali:2005fu}. Our model tries to mimic this behaviour using the following expression:
\begin{equation}
V_{\rm CON}(\vec{r}\,)=\left[-a_{c}(1-e^{-\mu_{c}r})+\Delta \right] (\vec{\lambda}_{q}^{c}\cdot\vec{\lambda}_{\bar{q}}^{c}) \,,
\label{eq:conf}
\end{equation}
where $a_{c}$ and $\mu_{c}$ are model parameters. At short distances this potential presents a linear behaviour with an effective confinement strength, $\sigma=-a_{c}\,\mu_{c}\,(\vec{\lambda}^{c}_{i}\cdot \vec{\lambda}^{c}_{j})$, while it becomes constant at large distances. This type of potential shows a threshold defined by
\begin{equation}
V_{\rm thr} = \{-a_{c}+\Delta\}(\vec{\lambda}^{c}_{i}\cdot \vec{\lambda}^{c}_{j}).
\end{equation}
No quark-antiquark bound states can be found for energies higher than this threshold. The system suffers a transition from a colour string configuration between two static colour sources into a pair of static mesons due to the breaking of the colour flux-tube and the most favoured subsequent decay into hadrons.

The OGE potential is generated from the vertex Lagrangian
\begin{equation}
{\mathcal L}_{qqg} = i\sqrt{4\pi\alpha_{s}} \, \bar{\psi} \gamma_{\mu} G^{\mu}_{c} \lambda^{c} \psi,
\label{Lqqg}
\end{equation}
where $\lambda^{c}$ are the $SU(3)$ colour matrices, $G^{\mu}_{c}$ is the gluon field and $\alpha_{s}$ is the strong coupling constant. The scale dependence of $\alpha_{s}$ can be found in {\it e.g.} Ref.~\cite{Vijande:2004he}, it allows a consistent description of light, strange and heavy mesons. 


Explicit expressions for all the potentials and the value of the model parameters can be found in Ref.~\cite{Vijande:2004he}, updated in Ref.~\cite{Segovia:2008zz}. Meson eigenenergies and eigenstates are obtained by solving the Schr\"odinger equation using the Gaussian Expansion Method~\cite{Hiyama:2003cu} which provides enough accuracy and it simplifies the subsequent evaluation of the needed matrix elements.

Following Ref.~\cite{Hiyama:2003cu}, we employ Gaussian trial functions with ranges in geometric progression. This enables the optimization of ranges employing a small number of free parameters. Moreover, the geometric progression is dense at short distances, so that it enables the description of the dynamics mediated by short range potentials. The fast damping of the Gaussian tail does not represent an issue, since we can choose the maximal range much larger than the hadronic size.


\subsection{Coupled-channel quark model}
\label{subsec:coupledchannel} 

The quark-antiquark bound state can be strongly influenced by nearby multiquark channels~\footnote{Note here that this effect is not trivial and, thus, not explicitly taken into account with the linear screened potential which considers an almost constant, global mass-shift of all naive meson states due to the presence of far meson-meson thresholds.}. In this work, we follow Ref.~\cite{Ortega:2009hj} to study this effect in the spectrum of the bottom-charmed mesons and thus we need to assume that the hadronic state is given by
\begin{equation} 
| \Psi \rangle = \sum_\alpha c_\alpha | \psi_\alpha \rangle + \sum_\beta \chi_\beta(P) |\phi_A \phi_B \beta \rangle,
\label{ec:funonda}
\end{equation}
where $|\psi_\alpha\rangle$ are bottom-charmed eigenstates of the two-body Hamiltonian, $\phi_{M}$ are wave functions associated with the $A$ and $B$ mesons, $|\phi_A \phi_B \beta \rangle$ is the two-meson state with $\beta$ quantum numbers coupled to total $J^{PC}$ quantum numbers and $\chi_\beta(P)$ is the relative wave function between the two mesons in the molecule. When we solve the four-body problem we use 
the $q\bar q$ wave functions obtained from the solution of the two-body problem using the Gaussian Expansion Method (GEM).

To derive the $B^{(\ast)}D^{(\ast)}$ interaction from the $q\bar{q}$ one described above we use the Resonating Group Method (RGM)~\cite{Tang:1978zz}. For the process $AB\to CD$ the direct potential ${}^{\rm RGM}V_{D}^{\alpha\alpha '}(\vec{P}',\vec{P})$, where no quarks are exchanged between mesons, can be written as
\begin{align}
&
{}^{\rm RGM}V_{D}^{\alpha\alpha '}(\vec{P}',\vec{P}) = \sum_{i,j} \int d\vec{p}_A\, d\vec{p}_B\, d\vec{p}_C\, d\vec{p}_D\, \times \nonumber \\
&
\times \phi_C^{\ast}(\vec{p}_C) \phi_D^{\ast}(\vec{p}_D) 
{\cal V}_{ij}^{\alpha\alpha '}(\vec{P}',\vec{P}) \phi_A(\vec{p}_{A}) \phi_B(\vec{p}_{B})  \,.
\label{eq:Kernel1}
\end{align}
where $\{i,j\}$ runs over the constituents of the involved mesons, $\alpha^{(\prime)}$ denotes the initial (final) channel quantum  numbers and ${\cal V}_{ij}^{\alpha\alpha'}$ is the CQM interaction between the $i$ and $j$ quarks (antiquarks).

Besides the direct potential, we can naturally connect meson-meson channels with different quark content with the quark rearrangement potential ${}^{\rm RGM}V_{R}^{\alpha\alpha'}(\vec{P}',\vec{P})$. This would allow us to study the decays to $B_c\pi$ channels from $B^{(\ast)}D^{(\ast)}$ states. The rearrangement potential is given by
\begin{align}
&
{}^{\rm RGM}V_{R}^{\alpha\alpha'}(\vec{P}',\vec{P}) = \sum_{i,j}\int d\vec{p}_A \,
d\vec{p}_B\, d\vec{p}_C\, d\vec{p}_D\, d\vec{P}^{\prime\prime}\, \phi_{A}^{\ast}(\vec{p}_C) \times \nonumber \\
&
\times  \phi_D^{\ast}(\vec{p}_D) 
{\cal V}_{ij}^{\alpha\alpha '}(\vec{P}',\vec{P}^{\prime\prime}) P_{mn} \left[\phi_A(\vec{p}_{A}) \phi_B(\vec{p}_{B}) \delta^{(3)}(\vec{P}-\vec{P}^{\prime\prime}) \right] \,,
\label{eq:Kernel2}
\end{align}
where $P_{mn}$ is the operator that exchanges quarks between clusters. 

The remaining part of our full interaction is the coupling between the quark-antiquark and meson-meson sectors which requires the creation of a light quark pair. The operator associated with this process should describe also the open-flavour meson strong decays and is given by~\cite{Segovia:2012cd}
\begin{equation}
\begin{split}
T =& -\sqrt{3} \, \sum_{\mu,\nu}\int d^{3}\!p_{\mu}d^{3}\!p_{\nu} \delta^{(3)}(\vec{p}_{\mu}+\vec{p}_{\nu})\frac{g_{s}}{2m_{\mu}}\sqrt{2^{5}\pi} \, \times \\
&
\times \left[\mathcal{Y}_{1}\left(\frac{\vec{p}_{\mu}-\vec{p}_{\nu}}{2} \right)\otimes\left(\frac{1}{2}\frac{1}{2}\right)1\right]_{0}a^{\dagger}_{\mu} (\vec{p}_{\mu})b^{\dagger}_{\nu}(\vec{p}_{\nu}) \,.
\label{eq:Otransition2}
\end{split}
\end{equation}
where $\mu$ $(\nu)$ are the spin, flavour and colour quantum numbers of the created quark (antiquark). The spin of the quark and antiquark is coupled to one. The ${\cal Y}_{lm}(\vec{p}\,)=p^{l}Y_{lm}(\hat{p})$ is the solid harmonic defined in function of the spherical harmonic. We fix the relation of $g_{s}$ with the dimensionless constant giving the strength of the quark-antiquark pair creation from the vacuum as $\gamma=g_{s}/2m$, being $m$ the mass of the created quark (antiquark).

It is important to emphasize here that the $^{3}P_{0}$ model depends only on one parameter, the strength $\gamma$ of the decay interaction. Some attempts have been done to find possible dependences of the vertex parameter $\gamma$, see~\cite{Ferretti:2013vua} and references therein. In Ref.~\cite{Segovia:2012cd} we performed a successful fit to the decay widths of the mesons which belong to charmed, charmed-strange, hidden charm and hidden bottom sectors and elucidated the dependence on the mass scale of the $^{3}P_{0}$ free parameter $\gamma$. Further details about the global fit can be found in Ref.~\cite{Segovia:2012cd}. The running of the strength $\gamma$ of the $^{3}P_{0}$ decay model is given by
\begin{equation}
\gamma(\mu) = \frac{\gamma_{0}}{\log\left(\frac{\mu}{\mu_{0}}\right)},
\label{eq:fitgamma}
\end{equation}
where $\gamma_{0}$ and $\mu_{0}$ are parameters, whereas $\mu$ is the reduced mass of the quark-antiquark in the decaying meson. The value of $\gamma$ that we are using in this work is the one corresponding to the bottom-charmed sector: $\gamma=0.247$. In order to quantify the sensitivity of the results with the coupling of the two sectors, we will explore a variation of $10\%$ in this parameter, thus we will use the range $\gamma=0.247\pm0.025$. The relative error coming from the fit in Ref.~\cite{Segovia:2012cd} is only of around $3\%$, however when one considers the average of relative errors of the states in Table III of Ref.~\cite{Segovia:2012cd} they are of the order of $13\%$ so, in order to be conservative, we take the latter value for the uncertainty.

From the operator in Eq.~(\ref{eq:Otransition2}), we define the transition potential $h_{\beta \alpha}(P)$ within the $^{3}P_{0}$ model as~\cite{Kalashnikova:2005ui} 
\begin{equation}
\langle \phi_{M_1} \phi_{M_2} \beta | T | \psi_\alpha \rangle = P \, h_{\beta \alpha}(P) \,\delta^{(3)}(\vec P_{\rm cm})\,,
\label{Vab}
\end{equation}
where $P$ is the relative momentum of the two-meson state.


Adding the coupling with bottom-charmed states we end-up with the coupled-channels equations
\begin{equation}
\begin{split}
&
c_\alpha M_\alpha +  \sum_\beta \int h_{\alpha\beta}(P) \chi_\beta(P)P^2 dP = E c_\alpha\,, \\
&
\sum_{\beta}\int H_{\beta'\beta}(P',P)\chi_{\beta}(P) P^2 dP + \\
&
\hspace{2.50cm} + \sum_\alpha h_{\beta'\alpha}(P') c_\alpha = E \chi_{\beta'}(P')\,,
\label{ec:Ec-Res}
\end{split}
\end{equation}
where $M_\alpha$ are the masses of the bare $c\bar{b}$ mesons and $H_{\beta'\beta}$ is the RGM Hamiltonian for the two-meson states obtained from the $q\bar{q}$ interaction. Solving the coupling with the bottom-charmed states, we arrive to a Schr\"odinger-type equation
\begin{equation}
\begin{split}
\sum_{\beta} \int \big( H_{\beta'\beta}(P',P) + & V^{\rm eff}_{\beta'\beta}(P',P) \big) \times \\
&
\times \chi_{\beta}(P) {P}^2 dP = E \chi_{\beta'}(P'),
\label{ec:Ec1}
\end{split}
\end{equation}
where
\begin{equation}\label{ec:effectiveV}
V^{\rm eff}_{\beta'\beta}(P',P;E)=\sum_{\alpha}\frac{h_{\beta'\alpha}(P') h_{\alpha\beta}(P)}{E-M_{\alpha}}.
\end{equation}

Finally, let us mention that this version of the coupled-channel quark model has been applied extensively to the study of XYZ states (see, for instance, Ref.~\cite{Ortega:2012rs}) and can describe both the renormalization of the bare $b\bar c$ states due to the presence of nearby meson-meson thresholds and the generation of new states through the meson-meson interaction due to the coupling with $b\bar c$ states and the underlying quark-antiquark interaction, as it is the case for the $X(3872)$~\cite{Ortega:2009hj} 
(see also~\cite{Tan:2019qwe} for a similar calculation but with some exploratory improvements of the 
coupling operator between meson and meson-meson sectors).


\begin{table}[!t]
\begin{center}
\caption{\label{tab:predmassesbc} Masses, in MeV, of $B_c$ states (with $n<=5$) predicted by our constituent quark model , compared to experiments and recent lattice QCD studies. }
\begin{tabular}{cccccccc}
\hline
\hline 
State & $J^{P}$ & $n$ & The.  & Ref.~\cite{Mathur:2018epb} & Ref.~\cite{Dowdall:2012ab} & Exp. & \\
\hline
$B_c$ & $0^-$ & $1$ & $6277$  & $6276\pm7$ & $6278\pm9$& $6274.9\pm0.8$ & \cite{Tanabashi:2018oca} \\
      &       & $2$ & $6868$  & -  &$6894\pm21$& $6871.0\pm1.7$ &\cite{Tanabashi:2018oca} \\
      &       & $3$ & $7248$ &- &- & - &\\
      &       & $4$ & $7534$ & -&- & - &\\
      &       & $5$ & $7761$ & -&- & - &\\[2ex]
$B_{c0}$ & $0^+$ & $1$ & $6689$ & $6712\pm19$ & $6707\pm16$ & - & \\
         &       & $2$ & $7109$ &- &- & - & \\
         &       & $3$ & $7421$ &- &- & - & \\
         &       & $4$ & $7668$ &- &- & - & \\
         &       & $5$ & $7868$ &- &- & - & \\[2ex]
$B_{c1}$ & $1^+$ & $1$ & $6723$ & $6736\pm18$ & $6742\pm16$ & - & \\
         &       & $2$ & $6731$ & - & - & - & \\
         &       & $3$ & $7135$ & - & - & - & \\
         &       & $4$ & $7142$ & - & - & - & \\
         &       & $5$ & $7442$ & - & - & - & \\
         &       & $6$ & $7449$ & - & - & - & \\[2ex]
$B_{c}^\ast$ & $1^-$ & $1$ & $6328$ & $6331\pm7$ & $6332\pm9$& - &\\
             &       & $2$ & $6898$ & - & $6922\pm21$ &  \\
             &       & $3$ & $6999$ & - & - & - & \\
             &       & $4$ & $7270$ & - & - & - & \\
             &       & $5$ & $7333$ & - & - & - & \\[2ex]
$B_{c2}$ & $2^+$ & $1$ & $6742$ & - & - & - & \\
         &       & $2$ & $7151$ & - & - & - & \\
         &       & $3$ & $7226$ & - & - & - & \\
         &       & $4$ & $7456$ & - & - & - & \\
         &       & $5$ & $7508$ & - & - & - & \\[2ex]
\hline
\hline
\end{tabular}
\end{center}
\end{table}

\section{RESULTS}
\label{sec:results}

Predictions of  our CQM for the low-lying $B_c$ states for $J^P=0^\pm,1^\pm$ and $2^+$  are shown in Table~\ref{tab:predmassesbc}. These are compared to the scarce experimental data from the PDG~\cite{Tanabashi:2018oca}. The recent results of CMS~\cite{Sirunyan:2019osb} and LHCb~\cite{PhysRevLett.122.232001} Collaborations coincides with our prediction within the experimental error in the $2^1S_0$ case. As stated in the introduction, the observed  $B_c^*(2S)$ peak is found at an energy lower than the one of the $B_c(2S)$. However this result should be taken with care because the low energy photon emitted in the $B_c^*\rightarrow B_c\gamma$ radiative decay is not reconstructed~\cite{Sirunyan:2019osb} and therefore we cannot compare this result with the theoretical one.

These experimental results only cover the lowest-lying states of the $0^-$ and $1^-$ sectors. To compare other sectors we included recent lattice QCD studies, such as the quenched $2+1$~\cite{Allison:2004be} and the $2+1+1$ flavors~\cite{Dowdall:2012ab} calculations of the HPQCD Collaboration, and the $2+1+1$ flavors analysis of Ref.~\cite{Mathur:2018epb}. An overall good agreement with the available lattice/experimental data for the $B_c$ spectra below the lowest $B^{(\ast)}D^{(\ast)}$ thresholds is obtained. 

Above those aforementioned thresholds coupled-channels effects may appear. The influence of the couplings of bare $q\bar q$ states with open channels depends on the relative position of the $q\bar q$ mass and the open threshold. One can see from Eq.~\eqref{ec:effectiveV} that when the value of the threshold energy $E$ is greater than the $q\bar q$ mass $M$ the effective potential is repulsive and it is unlikely that the coupling can generate a bound state. However if $M>E$ the potential becomes negative and an extra bound state with a large molecular probability may appear.

In analogy with charm-strange ($D_s$) mesons, where a rich phenomenology is found in the $0^+$ and $1^+$ sectors (see, e.g. Ref.~\cite{Ortega:2016mms}), it is interesting to evaluate the positive-parity sectors, at least for those in which $B^{(*)}D^{(*)}$ channels are in a relative S-wave. This allows significant couplings between both two- and four-quark sectors, which could produce deviations from quenched quark model calculations or produce new unexpected states. 

In order to evaluate the effect of the closest thresholds, we consider all the $B_c$ states predicted by CQM (Table~\ref{tab:predmassesbc}), within $\pm 150$ MeV around the closest open $D^{(\ast)} B^{(\ast)}$ threshold in $S$ or $D$ wave, whose masses are shown in Table~\ref{tab:thresholds}. 

 For the $J^P=0^+$ sector we study the $n=2$ and $3$ $^3P_0$ states coupled to the $DB$ molecule, for the $J^P=1^+$ sector we couple the $D B^\ast$ to the $3^3P_1$ and $4^1P_1$ whereas the $5^3P_1$ and $6^1P_1$ $1^+$ $B_c$ states are coupled to $D^\ast  B$ molecule. Finally, for the $J^P=2^+$ we couple the $D^\ast B^\ast$ to the $3^3F_2$ and $4^3P_2$ $B_c$ states.
The effect of further thresholds in the $B_c$ spectra is smooth and expected to be encoded in the screened confinement potential as a global contribution.

\begin{table}
 \caption{\label{tab:thresholds} Masses [MeV] of the isospin-averaged $D^{(\ast)} B^{(\ast)}$ thesholds, from PDG~\cite{Tanabashi:2018oca}.}
 \begin{tabular}{ccccc}
  \hline
  \hline
  Channel &  $D B$ & $D B^\ast$ & $D^\ast B$ & $D^\ast B^\ast$ \\
  \hline
  Mass    &  7146.57   & 7192.33  & 7287.96 &  7333.72 \\
  \hline
  \hline
 \end{tabular}
\end{table}

\begin{table}
 \caption{\label{tab:CQMbare} Theoretical bare $b\bar c$ masses (in MeV) within $\pm 150$ MeV from 
 $D^{(\ast)} B^{(\ast)}$ thesholds of Table~\ref{tab:thresholds}, selected for the coupled-channels calculation. 
 $\Delta M\equiv M_{M_{\rm thres}-\rm bare}$ (in MeV) shows the distance to closest 
 threshold (see Table~\ref{tab:thresholds}).}
 \begin{tabular}{ccrr}
  \hline
  \hline
  $J^{PC}$  & $n^{2S+1}L_J$ & Mass & $\Delta M$ \\
  \hline
 $0^{+}$   & $2^3P_0$ & 7109 & 37.84\\
           & $3^3P_0$ & 7421 & -87.54\\
 $1^{+}$   & $3^3P_1$ & 7135 & 57.47\\
           & $4^1P_1$ & 7142 & 50.13\\
           & $5^3P_1$ & 7442 & -108.47\\
           & $6^1P_1$ & 7449 & -115.44\\           
 $2^{+}$   & $3^3F_2$ & 7226 & 107.46 \\
           & $4^3P_2$ & 7456 & -122.24 \\
  \hline
  \hline
 \end{tabular}
\end{table}

\begin{table}
\caption{Additional and dressed $b\bar c$ states. ${\cal P}_{b\bar c}^{\max}$ denotes the probability of the dominant $b\bar c^{\rm max}$ state. Theoretical error estimated from the uncertainty of the vertex parameter: $\gamma=0.247\pm0.025$.\label{tab:Resbc}}
\begin{tabular}{lccccc}
\hline
\hline
$J^{PC}$ & Mass [MeV] & Width [MeV] & ${\cal P}_{\rm mol}$ [\%] & $b\bar c^{\rm max}$ & ${\cal P}_{b\bar c}^{\max}$ [\%]   \\
\hline
$0^+$ & $7198\pm 6$ & $64\pm5$ & $35\pm6$ & $2^3P_0$ & $65\pm6$ \\
      & $7420.96\pm 0.05$ & $0.5\pm0.1$ & $57\pm1$ & $3^3P_0$ & $43\pm1$ \\
$1^+$ & $7109\pm 4$ & $0$ & $14.2^{+1.5}_{-1.6}$ & $3^3P_1$ & $85.8^{+1.6}_{-1.5}$ \\
      & $7117^{+4}_{-5}$ & $0$ & $7.8\pm1.1$ & $4^1P_1$ & $92.1\pm1.1$\\
      & $7436\pm 1$ & $40.86^{+14}_{-10}$ & $40^{+2}_{-3}$ & $5^3P_1$ & $52^{+3}_{-2}$ \\
      & $7360^{+7}_{-5}$ & $40^{+11}_{-7}$ & $67^{+5}_{-6}$ & $6^1P_1$ & $19^{+5}_{-4}$ \\
$2^+$ & $7222.6^{+0.7}_{-0.8}$ & $0$ &  $1.2\pm0.2$ & $3^3F_2$ & $98.8\pm0.2$\\
      & $7333.68^{+0.04}_{-0.2}$ & $0$ & $99.5^{+0.5}_{-1.0}$ & $4^3P_2$ & $0.5^{+1.0}_{-0.5}$\\
      & $7401^{+8}_{-7}$ & $42\pm 5$ & $52.7\pm5$ & $4^3P_2$ & $47.0\pm5$\\
\hline
\hline
\end{tabular}
\end{table}

In Table~\ref{tab:Resbc} we show the results for the positive-parity $B_c$ states near thresholds. The interaction derived from RGM does not bind the $D^{(\ast)} B^{(\ast)}$ by itself, as it happened for some cases on the bottomonium sector~\cite{Entem:2016ojz}, so the coupling could be a relevant dynamical mechanism to generate new states. 

For the $0^+$ sector we have two different situations: the bare $3^3P_0$ $q\bar q$ state is above the $DB$ threshold whereas the $2^3P_0$ is below. However the attraction produced in the first case is not enough to generate a new state. The results is that the bare mass of both states is renormalized and the two states acquires a molecular component being more significant in the $3^3P_0$ case.

 The two bare $3^3P_1$ and $4^1P_1$ $1^{+}$ states are below the $D B^\ast$ threshold which produce a repulsive interaction. The two states are slightly renormalized, maintaining their mass splitting while acquiring a small molecular component. The other two states  of the $1^+$ sector ($5^3P_1$ and $6^1P_1$) are above the $D^\ast B$ threshold. However, as in the case of the $3^3P_10$ state, the generated attraction is not enough to produce new molecular states but both states acquire an important molecular component.

Finally in the $2^+$ sector we have again one state ($3^3F_2$) below threshold and the other ($4^3P_2$) above threshold. The first practically remains as a pure $q\bar q$ state. However in the case of the $4^3P_2$ state  the attraction generated by the coupling is strong enough  to produce, besides a renormalized $4^3P_2$ state  a mostly molecular shallow $D^*B^*$ extra state, which we call $X_{B_{c2}}$. As the appearance of this new state is due to the coupling 
between molecular and $q\bar q$ components, its mass depends on the value of the strength of the $^3P_0$
parameter $\gamma$. With the central value obtained in Ref.~\cite{Segovia:2012cd} we get a very
loosly bound state. Considering a possible deviation of the order of $10\%$ the state can also be
a virtual state very close to threshold. This pole structure should induce enhancements in reactions
strongly coupled to the $D^\ast B^\ast$ $2^+$ channel near threshold that could be measured in
future experiments.


\section{Summary}
\label{sec:epilogue}

In this work we have studied the influence of two meson thresholds on the $B_c$ states in the 
$J^P$ sectors $0^+$, $1^+$ and $2^+$. In analogy to the charmonium sector the $2P$ states of
the $q\bar q$ spectra gets dressed by the molecular components of closeby thresholds. In the
charmonium sector an additional state, the $X(3872)$, in the $J^{PC}=1^{++}$ channel appears.
For the $B_c$ states we also find an additional state in the $J^P=2^+$ channel very close
to the $D^\ast B^\ast$ threshold. This state does not appear if we do not include the coupling
with the $q\bar q$ components, exactly in the same way as happened for the $X(3872)$. 
Its bound-state nature cannot be clearly stated due to model uncertainties, but an enhancement in the $D^\ast B^\ast$ channel above threshold is expected due to the attractive nature of the total interaction, which could be experimentally measured.


\begin{acknowledgments}
%
%
This work has been partially funded by Spanish  Ministerio de Econom\'ia, Industria y  Competitividad under contracts no. FPA2017-86380-P and FPA2016-77177-C2-2-P and
by the EU STRONG-2020 project under the program
H2020-INFRAIA-2018-1, grant agreement no. 824093.
J.S. acknowledges the use of the computer facilities of C3UPO at the Universidad Pablo de Olavide.\end{acknowledgments}


\bibliography{Bcmol}

\end{document}